\documentclass[12pt,preprint]{aastex}

\begin{document}

\newcommand{\HDbright}{HD 95934}			
\newcommand{\HIP}{HIP 80460}				
\newcommand{\gammaray}{$\gamma$-ray}	


\title{STACEE Observations of Markarian 421 During an Extended Gamma-Ray Outburst}
\shorttitle{STACEE Observations of Mrk421 in 2001}

\author{
	L.~M.~Boone\altaffilmark{1},
	J.~A.~Hinton\altaffilmark{2,11},
	D.~Bramel\altaffilmark{3},
	E.~Chae\altaffilmark{2},
	C.~E.~Covault\altaffilmark{4},
	P.~Fortin\altaffilmark{5},
	D.~M.~Gingrich\altaffilmark{6,7},
	D.~S.~Hanna\altaffilmark{5},
	R.~Mukherjee\altaffilmark{3},
	C.~Mueller\altaffilmark{5},
	R.~A.~Ong\altaffilmark{8},
	K.~Ragan\altaffilmark{5},
	R.~A.~Scalzo\altaffilmark{2},
	D.~R.~Schuette\altaffilmark{8,9},
	C.~G.~Th\'eoret\altaffilmark{5,10},
and
	D.~A.~Williams\altaffilmark{1}
}
\altaffiltext{1}{Santa Cruz Institute for Particle Physics, University of 
		California, Santa Cruz, CA 95064, USA}
\altaffiltext{2}{Enrico Fermi Institute, University of Chicago, 5640 Ellis Ave.,
		Chicago, IL 60637, USA}
\altaffiltext{3}{Columbia University \& Barnard College, New York, NY 10027, USA}
\altaffiltext{4}{Department of Physics, Case Western Reserve
		University, 10900 Euclid Ave., Cleveland, OH 44106, USA}
\altaffiltext{5}{Department of Physics, McGill University, Montreal, Quebec H3A
		2T8, Canada}
\altaffiltext{6}{Centre for Subatomic Research, University of Alberta, Edmonton,
		Alberta T6G 2N5, Canada}
\altaffiltext{7}{TRIUMF, Vancouver, British Columbia V6T 2A3, Canada}
\altaffiltext{8}{Department of Physics \& Astronomy, University of California,
		Los Angeles, CA 90095, USA}
\altaffiltext{9}{Present address: Department of Physics, Cornell University,
		Ithaca, NY 14853}
\altaffiltext{10}{Present address: Laboratoire de Physique Corpusculaire et
		Cosmologie, Coll\`ege de France, F-75231 Paris CEDEX 05, France}
\altaffiltext{11}{Present address: Max-Planck-Institut f\"ur Kernphysik
		Postfach 10 39 80, D-69029 Heidelberg, Germany}


\begin{abstract}

The active galaxy Markarian 421 underwent a substantial outburst in early 2001.
Between January and May of that year, the STACEE detector was used to observe the
source in \gammaray s between the energies of 50 and 500 GeV.  These observations 
represent the lowest energy \gammaray\  detection of this outburst by a
ground-based experiment.  Here we present results from these observations, which
indicate an average integral \gammaray\  flux of
$(8.0\pm0.7\pm1.5)\times 10^{-10}$ ${\rm cm}^{-2}{\rm s}^{-1}$
above 140 GeV. We also present a light curve for Markarian 421 as observed by
STACEE from March to May, and compare our temporal, as well as spectral,
measurements to those of other experiments.
\end{abstract}



\section{Introduction}

The blazar Markarian 421 is one of only a handful of astrophysical objects
detected in the very high energy (VHE) \gammaray\  regime, between 10 GeV and 100
TeV.  At a red-shift of 0.031, it is the closest BL Lac object seen by EGRET, and
the first to be detected by a ground-based TeV instrument \citep{punch92,petry96}.
Along with Markarian 501 (z=0.034), it is one of the most prominent sources of
extra-galactic VHE \gammaray s, and has been regularly monitored by atmospheric
Cherenkov telescopes since its TeV identification
\citep[see][for a review]{catanese99}.

Markarian 421 is an X-ray selected BL Lac object, and as such, it exhibits the
double-humped spectrum that is characteristic of blazars
\citep[e.g.][]{takahashi00}. The lower energy hump, believed to arise from
synchrotron radiation, peaks in the X-ray range, while the higher energy hump,
generally attributed to inverse Compton (IC) scattering of soft photons, peaks in
the GeV to TeV energy range.

The details of the emission processes that produce the IC hump are, as yet,
unresolved. For example, the origin of the soft photons that seed the IC
component of the spectrum is an outstanding question in the study of blazars.
Although synchrotron self-Compton models, in which the seed photons originate
from synchrotron radiation within the blazar jets, are generally favored for
Markarian 421 \citep{coppi92,coppi99}, other competing models exist.  These
include external Compton models \citep{dermer92,sikora94}, proton-induced
cascades \citep{mannheim93}, and proton synchrotron models
\citep{aharonian00,mucke01}.

Markarian 421 has been given to strong periods of flaring activity over the past
few years.  Of note was a flare in 1996 during which the source was detected at
more than ten times the flux of the Crab Nebula \citep{gaidos96}.  In the early
part of 2001, a rather impressive flare was again observed \citep{iau}, with
reported TeV fluxes of comparable magnitude to those recorded in 1996, but of
much longer duration \citep{krennrich01}.

It was during the 2001 flare that STACEE--48, an intermediate incarnation of
STACEE (the Solar Tower Atmospheric Cherenkov Effect Experiment), was commencing
operation. STACEE was able to observe the activities of Markarian 421 for much of
the flaring period.  These observations are of particular note as they represent
the only \gammaray\  detection below 200 GeV during this flare.


\section{The STACEE--48 Detector}

STACEE is a low threshold ground-based \gammaray\  detector located at the
National Solar Thermal Test Facility near Albuquerque, New Mexico.  STACEE uses
the large steerable mirrors (heliostats) of an existing solar research facility
to collect Cherenkov photons from particle cascades in the atmosphere.  The
photons are focused, via secondary optics located on a central tower, onto an
array of photomultiplier tubes, and their arrival times are recorded and
processed by high-speed electronics. Because of their very large collection areas,
wavefront sampling detectors, such as STACEE and CELESTE \citep{denaurois02}, are
currently the only ground-based instruments capable of detecting \gammaray s
below 200 GeV.  STACEE was commissioned for its first science observations with
32 heliostats \citep[STACEE--32, see][]{hanna02} and was used to detect
\gammaray\  emission from the Crab Nebula in 1998 and 1999 \citep{oser01}.

STACEE--48 \citep{covault01} was an upgrade of the prior 32-channel experiment
in which 16 new heliostats were added, bringing the total to 48, each with a mirror
area of 37 m$^2$.  The introduction of these new heliostats had the dual advantage
of increasing the total collection area, as well as producing a more favorable
detector geometry.  In addition, by incorporating these new channels into the
trigger, the energy threshold of the experiment was lowered from 190 GeV to
140 GeV due to improved suppression of the night sky background.

The trigger hardware for STACEE--48 was re-instrumented with custom built modules
that handle the tasks of signal delay and trigger formation \citep{ragan00}.
The delays are constructed from a 125 MHz pipeline capable of storing and delaying
one hit per channel in each 8 ns bin.  The system has a coincidence resolution of
1 ns, by virtue of a vernier encoding scheme, and a maximum possible delay of 2048
ns.  It applies delays to each channel such that an idealized Cherenkov event will
produce a coincident signal across the entire detector.  Channels are grouped by
geography into subclusters of eight heliostats each, and a trigger was formed if
four of the six subclusters were hit within a 28 ns window.  Individual subclusters
registered a hit if five of their eight constituent channels fired within a 12 ns
window.

Improvements from STACEE--32 were also made on a number of other fronts.  We
aligned the mirror facets on all 48 heliostats using a laser look-back system.
The phototubes were re-calibrated, and then re-distributed in the camera to
better balance the detector response.  We used the full Moon to align the optics
from end to end, which resulted in important improvements to the optical
throughput.  We also reduced the background light reflected from the asphalt on
the heliostat field by up to 40\% by resurfacing the field with a dark sealant.

We have conducted extensive simulations of the STACEE--48 detector and its
response to both \gammaray\  and hadronic events.  We employ the CORSIKA
air-shower simulation package \citep{corsika} with the QGSJET hadronic interaction
model \citep{qgsjet}, together with custom optical and electronics simulation
packages to obtain the detector response.  We have been successful in reproducing
various detector characteristics with the simulations, including the detector
rate at zenith and the trigger rate as a function of discriminator threshold. From
simulations of our effective area convolved with a \gammaray\  spectrum of
differential index $\alpha=2.1$ (see Fig.~\ref{fig-convolve}), the energy
threshold for our observations (defined as the peak of the detected differential
spectrum) was determined to be $140\pm 20$ GeV.  The error
represents the shift in the peak location as the spectral index is varied between
$1.7$ and $2.5$.


\section{Observations}

STACEE--48 was used to observe Markarian 421 in the early part of 2001
\citep{boone02}.  STACEE uses an on-off technique for source observations as
described in \citet{oser01}.  We take on-source data for 28 minutes, with a
corresponding off-source run of 28 minutes to form an on-off pair. Our observations
from March to May included about 26 nights, yielding 78 on-off pairs from 36 hours
of source observations and an equal amount of background observations.

We have found that the subcluster firing rates, which are driven primarily by the
night-sky background, are a powerful diagnostic of abnormal fluctuations in sky
conditions.   In good-quality pairs, the time evolution of the subcluster rates
for both on and off-source runs are very linear, and well correlated.  However,
in runs with passing clouds or high haze, the rates are irregular and
uncorrelated.  Thus, removing runs whose subcluster rates exhibit deviations from
a linear time-evolution has proven to be an effective quality selection criterion.
Of the original 78 pairs, 17 were removed from the data set using this criterion.

Three on-off pairs were removed because they showed abnormally high rates from
background light, and two pairs were removed due to heliostat malfunctions. Other
pairs were trimmed to remove short-term contamination from an otherwise
acceptable run.  Removing and trimming runs reduced the usable data to 56 on-off
pairs, totaling 22.0 hours of observations.

After applying all the selection criteria, the data set consists of $317,045\pm563$
on-source events and $307,641\pm555$ off-source events, yielding a $9,404\pm790$
event excess in the on-source data.  This corresponds to a raw average excess of
$7.1\pm0.6$ events/min, where the quoted error is statistical only.

Of particular concern to non-imaging atmospheric Cherenkov detectors like STACEE
is the relatively bright star \HDbright\ (magnitude 6.16 in the B band), which is
within a few arc-minutes of Markarian 421, and well within the STACEE field of
view.  A calculation of the expected photon flux from this star and the trigger
multiplicity required by the detector indicates that we should see much less than
one false trigger per night from the star alone.  However, random fluctuations in
the excess noise generated on each channel by the starlight may promote an
otherwise sub-threshold cosmic-ray event.  Such promotions of sub-threshold
background events would only occur in on-source observations, as there is no
corresponding star in the off-source field of view.  In this way, it may be
possible to create a fake signal. 

To study this star effect, we spent some time observing \HIP\
(RA=16.42 hr, Dec=37.39$^\circ$), a star with a
magnitude and declination comparable to \HDbright.  We took 12 on-off pairs,
consisting of approximately five
on-source hours.  After applying analogous quality requirements to those described
above, 11 pairs and 4.3 hours of data survive.  These data indicate a raw excess of
$1.9\pm1.2$ events/min from the star effect.

After correcting for this fake signal, the detector deadtime, and the distributions
of detector configurations (as described in the next section), we detect an average
\gammaray\  rate from Markarian 421 of $7.7\pm0.7\pm1.2$ $\gamma$/min, where the
first error is
the statistical error on the Markarian 421 measurement alone, and the second represents
the statistical error in the observations of \HIP, used to correct for the effect
of the star \HDbright.


\section{Flux Determination}\label{sec-spectrum}

Although the source is expected to be highly variable, it is useful to calculate
an average flux over all the STACEE--48 observations.  Detector simulations
indicate that the trigger rate varies significantly as a function of zenith angle,
and, to a lesser extent, detector thresholds.  In order to calculate quantities
that are relatively independent of these effects, we corrected the measured rate
at a given detector zenith angle and threshold to the corresponding expected rate
at a fiducial configuration (a zenith angle of $5^\circ$, near Markarian 421
transit, and a detector threshold of $\sim 5$ photoelectrons on each phototube).
To do this, we simulated the detector response to an assumed source spectrum at
the fiducial configuration, as well as the observation configuration.  We then
scaled the measured rate by the ratio of these responses to obtain a corrected
constant-configuration curve.

These corrections are sensitive to assumptions about the functional form of the
incident \gammaray\  flux.  Thus, we must first choose a hypothesis for what the
\gammaray\  spectrum might be in our energy range.  Figure \ref{fig-flux2} is a
spectral energy distribution of Markarian 421 showing measurements by EGRET,
averaged over five years of observations \citep{egret3}, and Whipple during the
2001 flare \citep{krennrich02}.  The STACEE butterfly was obtained by assuming a
simple power-law spectrum with a given spectral index. The normalization was then
fit such that the convolution of the assumed spectrum and the detector effective
area matched the actual rate measured by STACEE.  This procedure was done for a
range of differential spectral indices from $2.00$ to $2.20$, yielding a derived
differential photon flux of 
$(6.4\pm0.5_{stat}\pm1.3_{sys})\times 10^{-9}{\rm cm}^{-2}{\rm s}^{-1}{\rm TeV}^{-1}$
at 140 GeV.  Here the statistical error is the error on the Markarian 421
measurement. The systematic error is obtained by adding in quadrature the
statistical error from the \HIP\  measurement ($1.0\times10^{-9}$), the variation
of the result with respect to the spectral index ($0.5\times10^{-9}$), and
$0.6\times10^{-9}$ from the uncertainty in the detector corrections
\citep[see][for details]{boone02}.  Note that this result represents an average
for the 2001 flare.  Though spectral variability was reported at higher energies
\citep{krennrich02,aharonian02}, STACEE--48 was not sufficiently sensitive to
resolve this variation.

Figure \ref{fig-fluxflux} is a plot of the correlation between the STACEE integral
flux above 140 GeV and corresponding data taken by the Whipple \gammaray\
telescope (Mt.~Hopkins, AZ, USA) above 390 GeV \citep{holder01}.
Points represent data from both
experiments whose time stamps agree to within 0.02 days ($\sim$29 minutes).  A linear
fit to the data yields a flux ratio of $4.3\pm 1.4$.
Assuming a simple power
law spectrum for
Markarian 421, this ratio corresponds to a differential spectral index of
$\alpha = 2.4^{+ 0.3}_{- 0.4}$.  If we include an exponential roll-off at
$E_o=4.3$ TeV, the derived index drops to
$\alpha = 2.3^{+ 0.3}_{- 0.4}$, both of which are consistent with the value
of $\alpha=2.14$ from \citet{krennrich01}.

Assuming a differential spectrum of index $\alpha=2.1$ over STACEE's energy range,
and after correcting for variations in the detector configuration, as well as for
the observed star effect,  we find the integral flux above 140 GeV to be
$(8.0\pm 0.7_{stat}\pm 1.5_{sys})\times 10^{-10}$ ${\rm cm}^{-2}{\rm s}^{-1}$.
The first error is the statistical error in the Markarian 421 observations, and
the second reflects the systematic errors discussed above, of which the error in
the star correction
is the dominant contribution ($\pm1.3\times10^{-10}$).


\section{Temporal Analysis}

While there is a substantial excess from Markarian 421 in the entire data set,
individual on-off pairs generally do not exhibit highly significant detections.
By combining the observations of a single night, we can improve the flux
significance at the cost of some temporal resolution.  To obtain the STACEE--48
light curve for Markarian 421, we averaged the rate over all the observations of
each night, and assigned a date which was a weighted average of the Modified Julian
Date in the middle of each run.

Figure \ref{fig-lightcurve} includes the STACEE--48 light-curve for Markarian 421
between early March and late May.  Filled circles are the counts per
minute, and error bars are statistical only.  For comparison, contemporaneous RXTE
quick-look results (provided by the RXTE All Sky Monitor team) are plotted above,
and Whipple observations \citep{holder01} are plotted below.  All the experiments
observe an increased flux in the latter part of March, falling to lower values in
April.  Note that the coincident gaps in both the STACEE and Whipple coverage
correspond to dates near the full Moon, during which Cherenkov telescopes generally
cannot operate.

The STACEE fluxes appear correlated with both the RXTE and Whipple observations.
In fact, a discrete correlation analysis \citep{krolik88} is consistent with
the STACEE and Whipple observations being correlated on day timescales.  However,
the limited statistics in the nightly STACEE observations currently make analysis
on shorter timescales infeasible.


\section{Conclusions}

STACEE has detected the BL Lac object, Markarian 421, with high significance in
the 140 GeV energy band, a hitherto unexplored region of its
spectral energy distribution.  The STACEE observed flux is consistent with
other VHE \gammaray\  observations, and the temporal evolution
of the STACEE observations appears similar to observations in both the TeV
and X-ray bands.

Figure \ref{fig-flux2} indicates that the STACEE measurements occupy an important
region in Markarian 421's spectral energy distribution near the peak of the IC
hump.  Future measurements of the spectrum by STACEE, along with simultaneous
data at X-ray and TeV energies should help to further constrain synchrotron and
IC blazar models for this source.  STACEE is currently operating with 64 channels,
each equipped with a flash ADC, and should be able to obtain more detailed spectral
information in the near future.


\acknowledgments

We are grateful to the staff of the National Solar Thermal Test Facility for their
excellent support.  Thanks to Gora Mohanty, Jeff Zweerink, Tumay Tumer, Marta
Lewandowska, Scott Oser, Fran\c{c}ois Vincent, and Joseph Boone.
This work was supported in part by the National Science Foundation (under Grant
Numbers
PHY-9983836,		
PHY-0070927,		
and PHY-0296052),	
the Natural Sciences and Engineering Research Council,
FCAR (Fonds pour la Formation de Chercheurs et l'Aide \`a la Recherche),
the Research Corporation and the California Space Institute.



\clearpage

\plotone{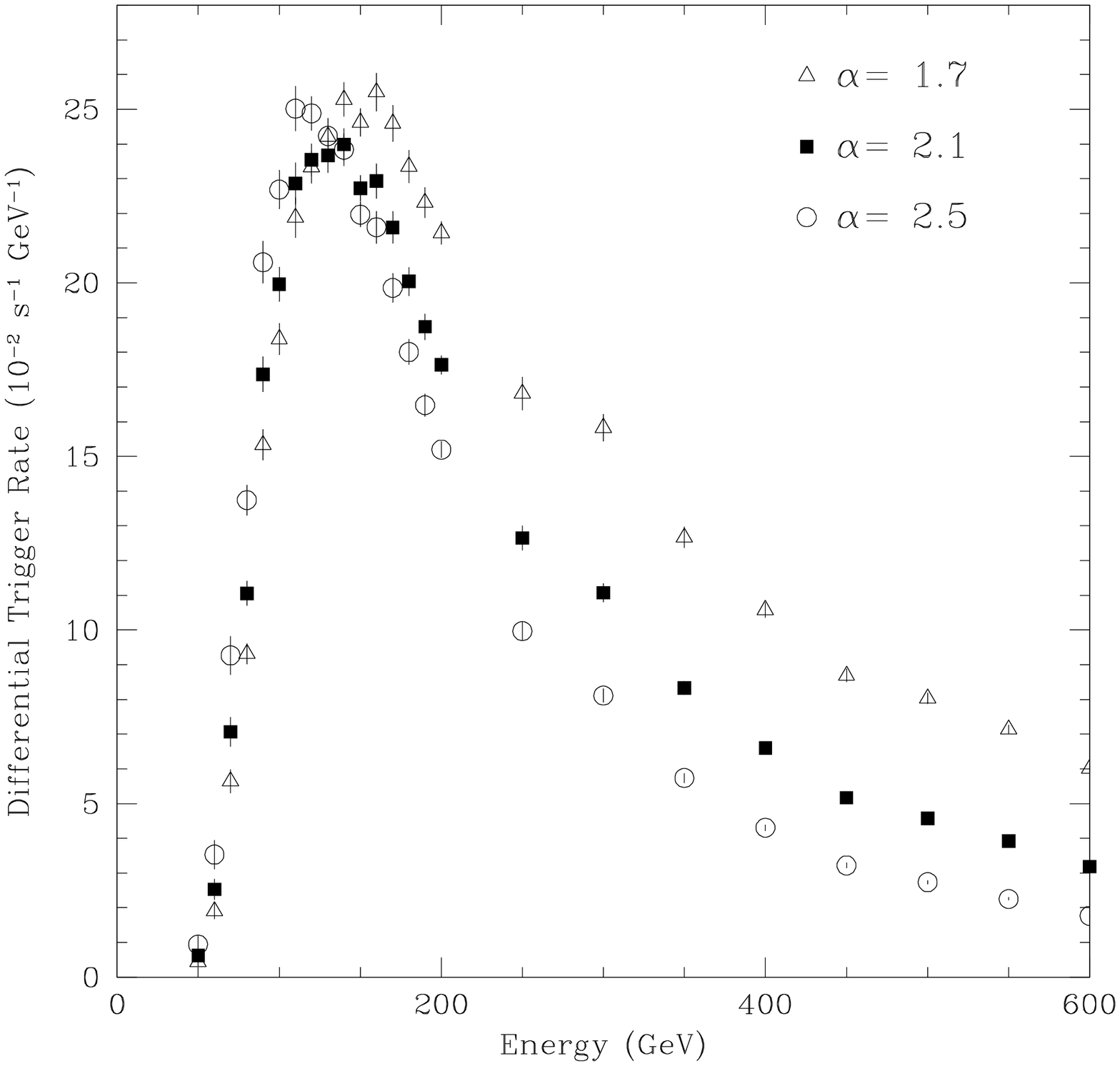}
\figcaption[stacee_difftrig_spectra.eps]{STACEE--48 detector response to several
different power-law spectra.  Spectra are normalized to the same incident
integral flux above 50 GeV for the given differential spectral index, $\alpha$.
\label{fig-convolve}}

\plotone{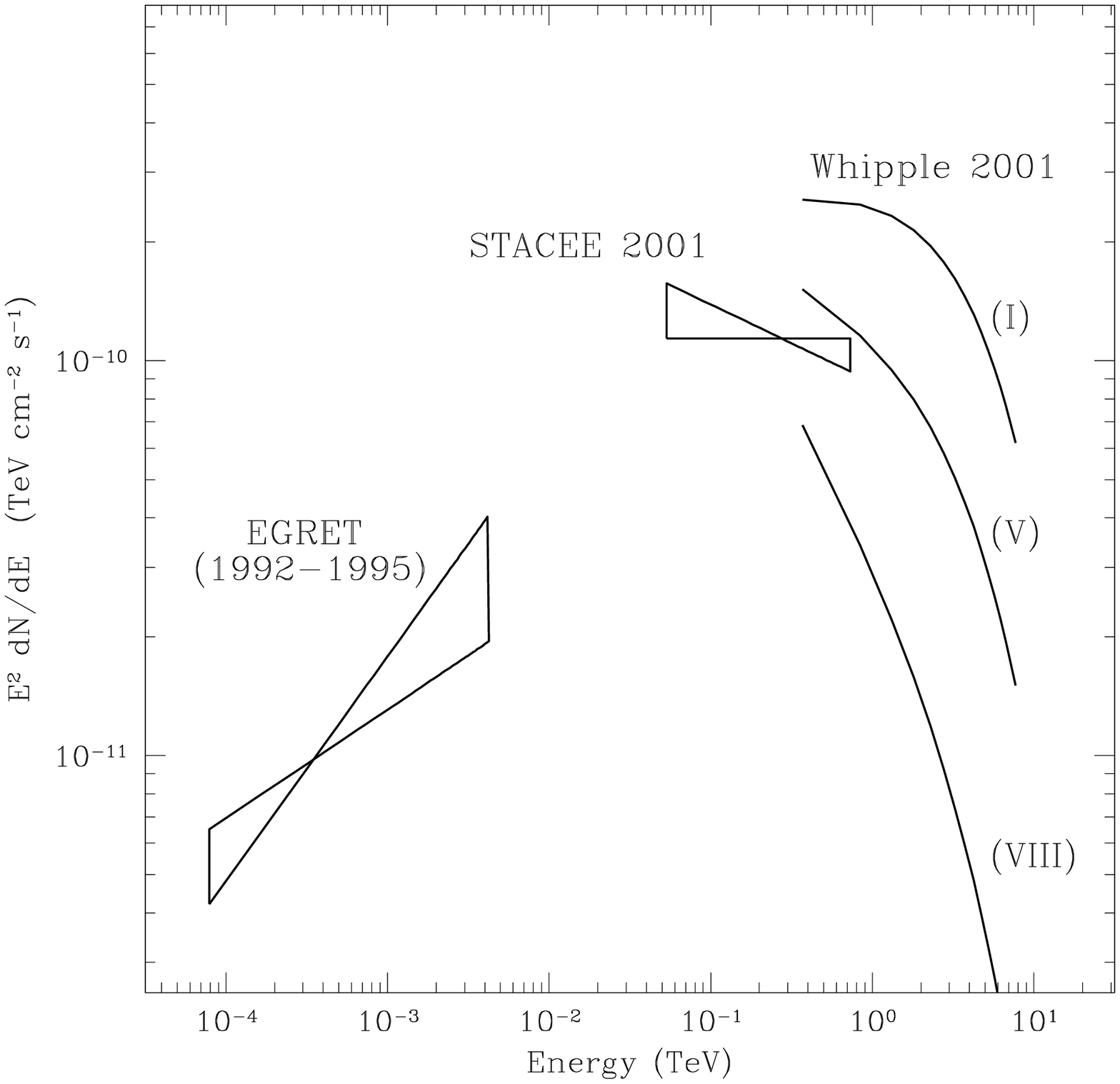}
\figcaption[diffflux.eps]{High-energy spectrum for Markarian 421.  The STACEE
butterfly assumes differential spectral indices between $\alpha$ = 2.00 and 2.20.
Note that, though the STACEE sensitivity peaks at 140 GeV, the vertex of the
butterfly is much higher.  STACEE data are averaged over the flaring period,
while the Whipple curves represent high (I), medium (V), and low (VIII) flux
levels over the course of the flare \citep[see][for details]{krennrich02}.  EGRET
data are from cycles 1 through 5, and thus represent an average low state.
\label{fig-flux2}}

\plotone{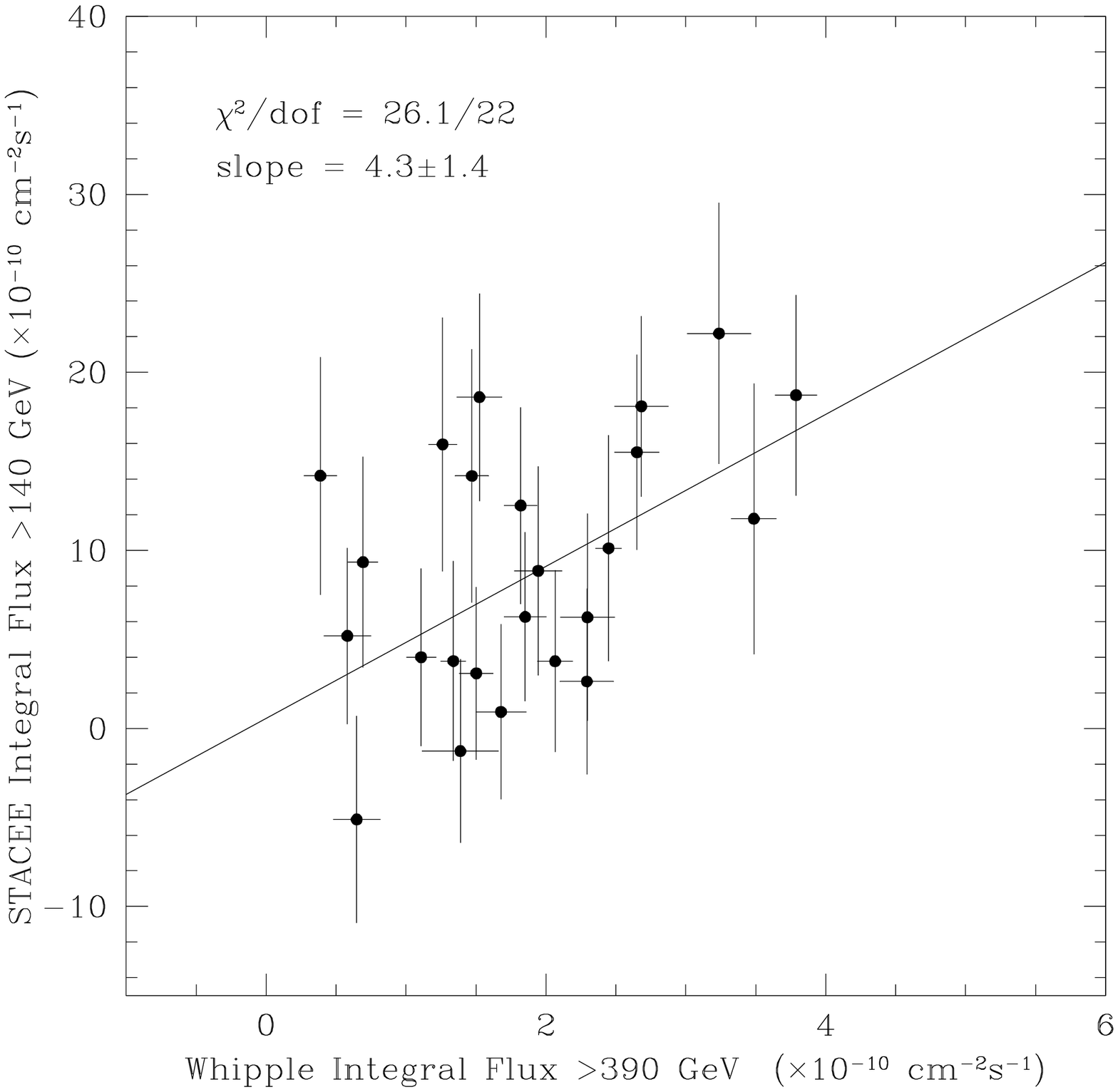}
\figcaption[fluxfluxD.eps]{Correlations between contemporaneous STACEE and Whipple
observations from March to May of 2001.  Points represent STACEE and Whipple runs
whose time stamps match to within 0.02 days (29 minutes).  The $\chi^2$/dof for
the fit is 26.1/22.  The slope of 4.27 is consistent with a differential spectral
index of $\alpha=2.14$, given the STACEE and Whipple thresholds.
\label{fig-fluxflux}}

\plotone{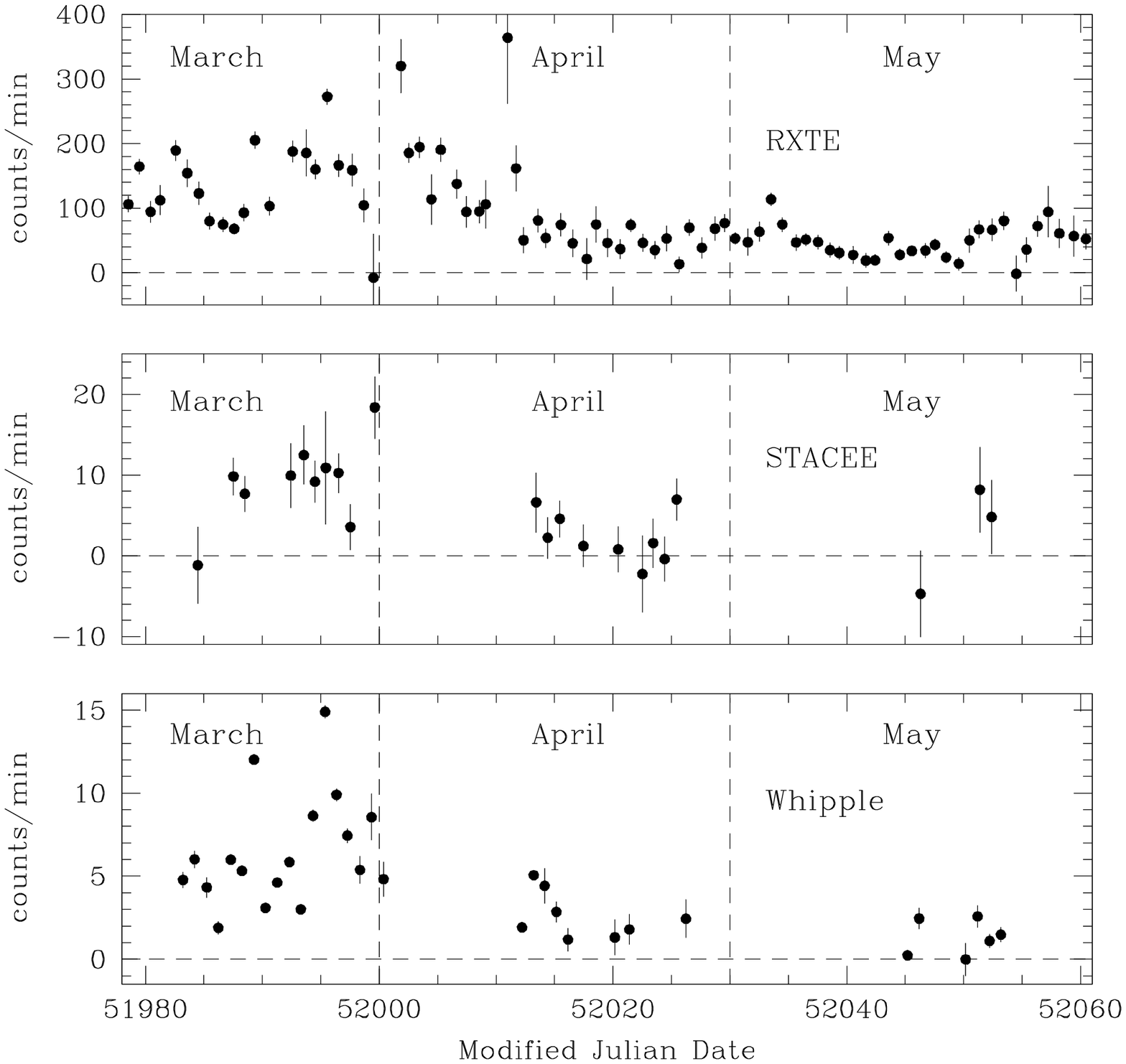}
\figcaption[multi.eps]{Markarian 421 light curve for Spring, 2001.  RXTE
observations are from the 2-10 keV band, STACEE data are from the 50-300 GeV band,
and Whipple observations span the 0.25-8 TeV band.  References cited in the
text.\label{fig-lightcurve}}

\end{document}